\documentclass[twocolumn,showpacs,preprintnumbers,amsmath,amssymb]{revtex4}
\usepackage{epsfig}
\preprint{NaCoO/2005}
\begin{document}
\title{Optical conductivity and charge ordering in Na$_{x}$CoO$_2$}
\author{S. Lupi, M. Ortolani, L. Baldassarre, P. Calvani}
\affiliation{"Coherentia" - INFM and Dipartimento di Fisica, Universit\`a di Roma La Sapienza, Piazzale Aldo Moro 2, I-00185 Roma, Italy}
\author{D. Prabhakaran, A. T. Boothroyd}
\affiliation{ Department of Physics, Oxford University, Oxford, OX1 3PU, United Kingdom}
\date{\today}

\begin{abstract}
The infrared conductivity $\sigma (\omega)$ of Na$_{x}$CoO$_2$ is studied as a function of doping and temperature for 0.5 $\le x \le$ 1. Charge localization in CoO$_2$ layers shows up through a far-infrared peak (FIP) in $\sigma (\omega)$, which coexists with a small Drude contribution. Long-range ordering at $x$ = 0.5 is confirmed to create a far-infrared gap, in addition to the FIP. At high $x$, the formation of a Spin-Density Wave reported below 22 K dramatically shifts the FIP to higher energy when $x$ is incommensurate with the lattice, indicating an abrupt deepening of the localizing potential. The in-plane E$_{1u}$ phonon lifetime is shown to be sensitive to both "freezing" and ordering of the mobile Na$^{+}$ ions. A comparison with the behavior of the FIP shows that such "freezing" is not the only origin of charge localization in the CoO$_2$ layers.
\end{abstract}
\pacs{74.25.Gz, 74.72.-h, 74.25.Kc}
\maketitle

\section{Introduction}

In low-dimensional, correlated electron systems, Coulomb, magnetic, and charge-lattice interactions are known to produce competitive ground states. For example, in cuprates at low temperature one may observe either unconventional superconductivity, or charge/spin ordering, or both. The recent discovery\cite{Takada} of the "wet" superconductor Na$_{x}$CoO$_2$$\cdot y$H$_2$O added a new oxide to the list of those which exhibit such intriguing properties.

The anhydrous parent compound of the superconducting cobaltates, Na$_{x}$CoO$_2$ (NCO), has a hexagonal crystal structure (P6$_3$/mmc). Its $ab$ planes are formed by CoO$_2$ layers separated by Na$^+$ ions. Such ions are highly mobile among three different lattice sites: in Na(1) the Na$^+$ ion is placed at (0,0,1/4), above the Co site of the CoO$_2$ layers; in Na(2) it occupies the (2/3, 1/3, 1/4) lattice point above the center of the triangular lattice of the CoO$_2$ layers; the third site Na(2)' is similar to the Na(2) one but Na$^+$ is at (2x, x, 1/4). The occupation probability of the above sites depends on both $x$ and $T$.
For $x$ = 0.5 the Na$^+$ ions are already arranged at room temperature in a periodic structure, both along the $ab$ plane and the $c$ axis, with Na(1) and Na(2) equally likely. At low $T$, Na$_{0.5}$CoO$_2$ becomes a charge-order insulator. It has been suggested that, since the Na$^+$ ions exert a strong unscreened potential on the CoO$_2$ layers, in the $x$ = 0.5 samples the charge-ordered state in the layers mirrors\cite{Zand,Huang} the periodic structure of the out-of-plane Na$^+$ ions. Na$^+$ order has been observed also for $x \neq$ 0.5 and seems more stable for commensurate doping ($x$ =  1/3; 1/2; 2/3; 3/4; 1).\cite{Huang,Bernhard};

The competition between charge mobility, localization, and ordering, in addition to a doping-dependent magnetic behavior, make the ($x,T$) phase diagram\cite{Foo} of Na$_{x}$CoO$_2$ nearly as rich as that of La-Ca manganites.\cite{Cheong}  Theory\cite{Lee} shows that CoO$_2$  is a Mott-Hubbard insulator, whereas NaCoO$_2$  should be a narrow-gap band insulator or a semiconductor. By increasingly adding Na$^+$ vacancies from $x = 1$, the electron ground state changes from a Spin-Density-Wave (SDW) metal (for 0.75 $\leq x < $ 1) to a Curie-Weiss metal (for $x <$ 0.75). For $x <$ 0.5 the metallic state is characterized by a more conventional Pauli paramagnetism. The two regions are separated by the above described 0.5 system. Once intercalated with water molecules, NCO becomes superconducting below about 5 K for 1/4 $< x <$ 1/3. The critical temperature $T_c$\cite{Zand,Milne} of Na$_{x}$CoO$_2$$\cdot y$H$_2$O depends on $y$, also because the intercalated  H$_3$O$^+$ ions are likely to provide additional doping to the CoO$_2$ planes.\cite{Milne}

The low-energy excitations of the CoO$_2$ plane have been studied by Angle-resolved Photoemission (ARPES) and infrared spectroscopy. ARPES detected, at $x$ = 0.7, a large hexagonal, hole-type, Fermi surface and a strongly renormalized quasi-particle band. This latter crosses the Fermi level in the \textbf{k}-space direction from $M$ to $\Gamma$. This corresponds to a single-particle hopping rate of 8 meV, consistently with the absence of quasi-free particles at high $T$.\cite{Hasan}

Infrared spectra of Na$_{x}$CoO$_2$ have been reported by several groups for different $x$.\cite{Lupi,Caimi,Wang,Hwang} Strong electron-phonon interaction and an anomalous Drude behavior have been found in $x$ = 0.57 single crystals, where the charge carriers have an effective mass of about 5 electron masses and their scattering rate $\Gamma(\omega)$  shows an $\omega^{3/2}$ dependence.\cite{Lupi} Similar results have been found for $x$ = 0.7, a sample close to a SDW instability.\cite{Caimi}  At $x$ = 0.5, the charge order shows up in the optical conductivity $\sigma (\omega)$ through the opening of a gap at low temperature around 150 cm$^{-1}$.\cite{Wang,Hwang} Meanwhile, a Far-Infrared Peak (FIP) springs up at 200 cm$^{-1}$, through a transfer of spectral weight ($SW$) across the gap.

However, an extended infrared study of charge localization and ordering in the (x,T) phase diagram of NCO has not been published yet.
Here we study the in-plane infrared conductivity $\sigma (\omega)$ of six single crystals of Na$_x$CoO$_2$  which cover the range 0.5 $\le x \le$ 1.0. In this whole range, the carriers are shown to be close to strong instabilities leading to localization and ordering. These phenomena show up through a strong FIP which at 0.5 is associated with a far-infrared gap. For $x \ge$ 0.75 the FIP is softer and coexists with a weak Drude term, while no gap is detected. A dramatic shift of the FIP to higher energies is observed, at high incommensurate $x$, between 30 and 12 K. This should be the effect of the spin-density-wave transition reported to occur in similar samples at 22 K.\cite{Foo} Finally, by studying the phonon line width, we also obtain interesting information on the relation between  charge localization in the CoO$_2$ planes and "freezing" of the out-of-plane Na$^+$ ions.

\section{Experiment and results}

Single crystals of Na$_x$CoO$_2$ (0.5 $\le x \le $ 1.0) have been grown by the floating-zone method in a four-mirror image furnace under an oxygen/argon atmosphere at 106 Pa.\cite{Prabha}
The reflectivity $R(\omega)$ of the six single crystals (x = 0.5, 0.57, 0.75, 0.85, 0.95, and 1.0) has been measured at quasi-normal incidence (8$^0$) between 320 and 8 K for 40 $< \omega <$ 20000 cm$^{-1}$, by using a Michelson interferometer and with radiation polarized along the CoO$_2$ layers. The final alignment of the reflectivity set-up has been performed with the interferometer under vacuum by using remotely controlled motors, to recover the mechanical stress of the optics due to interferometer evacuation. The crystals have been cleaved before each measurement to obtain a fresh and shiny surface. The reference was obtained by evaporating a gold or silver film, depending on the frequency range, onto the sample via a hot filament placed in front of it.

\begin{figure}
{\hbox{\psfig{figure=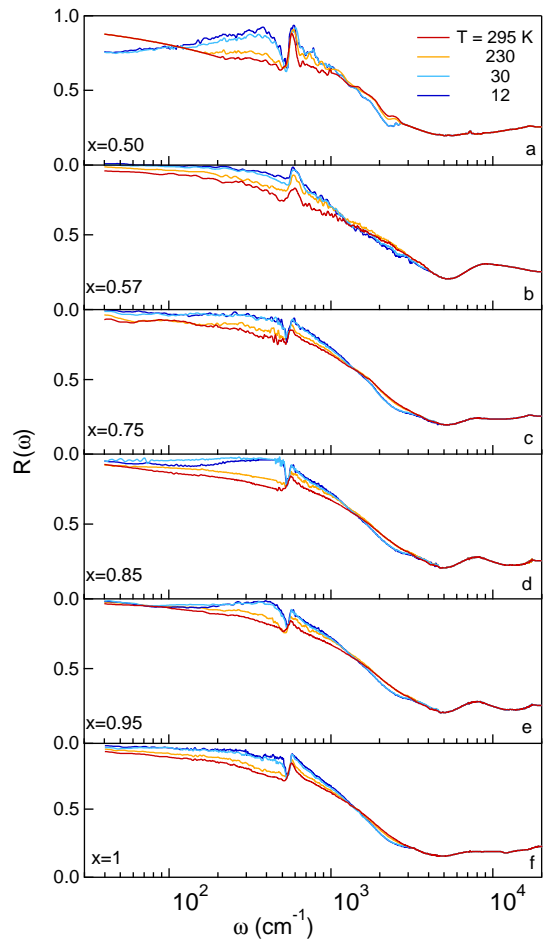,width=8cm}}} \caption{In-plane
infrared reflectivity of Na$_x$CoO$_2$ for 0.5 $\le x \le$ 1, at
different temperatures. For the sample with $x$ = 0.57, the
lowest two temperatures are 90 and 20 K.} \label{R}
\end{figure}

The reflectivity is shown at different temperatures in Fig.\ \ref{R}. That of the $x$= 0.57 sample has been already published.\cite{Lupi}  At room temperature, $R(\omega)$ shows,  for any Na content, an overall metal-like behavior,  indicated by its increase for decreasing frequencies. A pseudo-plasma edge is also evident around 5000 cm$^{-1}$, for any $x$ but 0.5, independently of the nominal doping. A similar effect has been observed in high-Tc cuprates. Therein, the pseudo-plasma edge could hardly be attributed to free-carrier absorption, like for the plasma edge of conventional metals. It was instead explained by an additional term centered in the mid-infrared, having both peak frequency and intensity roughly independent of doping.\cite{Calvani} At $x$ = 0.5, on the other hand, (Fig.\ \ref{R}-a), the pseudo-plasma edge is smeared compared to the other samples. Correspondingly, a lower value of the reflectivity is observed in the far- and mid-infrared range at all temperatures.

As $T$ decreases, $R(\omega)$ monotonously increases below 1500 cm$^{-1}$ in the $x$ = 0.57, 0.75, and 1 samples (Fig.\ \ref{R}-b, -c, and -f, respectively),  through a transfer of spectral weight across that frequency. In the 0.5, 0.85, and 0.95 crystals instead (Figs.\ \ref{R}-a, -d, and -e, respectively), high-frequency spectral weight is transferred mainly to a bump around 400 cm$^{-1}$. A shallow minimum appears in $R(\omega)$ around 150 cm$^{-1}$. At lower frequencies the reflectivity increases again to 1.
All samples show a strong phonon absorption around 590 cm$^{-1}$. This corresponds to a "layer-sliding" E$_{1u}$ mode related to the hexagonal P6$_3$/mmc structure, which results from the in-plane Co-O stretching motion mixed with a small component of the Na vibration parallel to the plane.\cite{Li} The other infrared-active phonon mode, predicted at about 200 cm$^{-1}$, is not observed. At high energy, the band peaked at 8000 cm$^{-1}$ has been attributed to a charge-transfer transition between the 2p-O states and the 3d-Co electronic states, while that around 15000 cm$^{-1}$ to a 3d-3d transition activated by a weak hybridization between 3d-Co and 2p-O states. \cite{Lupi,Caimi,Wang}

\begin{figure}
{\hbox{\psfig{figure=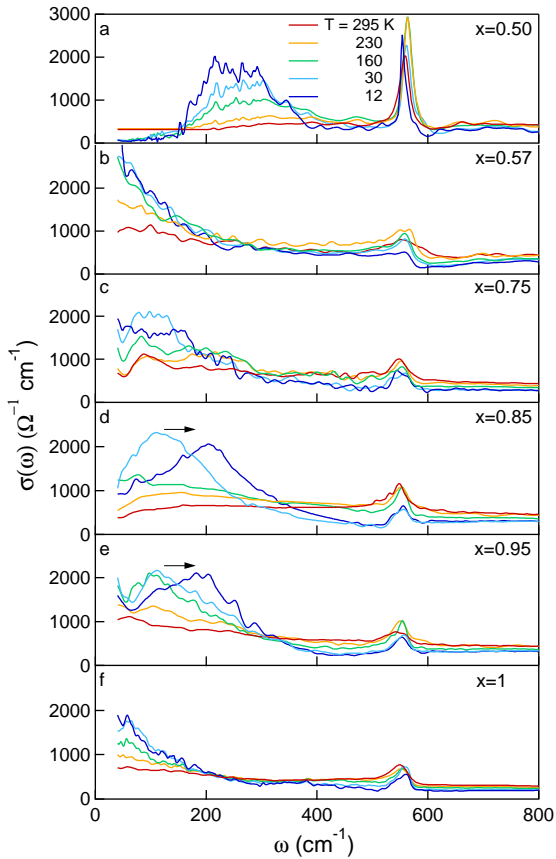,width=8cm}}} \caption{Real part
of the infrared conductivity of Na$_x$CoO$_2$ for 0.5 $\le x \le$
1, at different temperatures. For the sample with $x$ = 0.57, the
lowest two temperatures are 90 and 20 K. The arrows mark the
strong displacement below 30 K of the FIP in both samples with
high and incommensurate doping ($x$ = 0.85, 0.95).} \label{sigma}
\end{figure}

The optical conductivity $\sigma (\omega)$ was then obtained from $R(\omega)$ through standard Kramers-Kronig transformations.  $R(\omega)$ was extrapolated to zero frequency by using a standard Hagen-Rubens behavior and, at high frequencies, by a Lorentzian fit. An alternative high-frequency extrapolation based on the reflectivity\cite{Terasaki} of  NdCoO$_3$  gave the same results within errors.
The results are shown in Fig.\ \ref{sigma} for all samples, at selected temperatures. It exhibits a strong temperature dependence up to 800 cm$^{-1}$, while its low-frequency behavior is certainly non-monotonouos vs. $x$.
At $x$ = 0.50 (Fig.\ \ref{sigma}-a) an over-damped FIP is centered around 400 cm$^{-1}$ both at 295 and 230 K. Below that frequency a flat background is present, indicating incoherent charge transport at high temperature in the CoO$_2$ layers. The peak increases in intensity by decreasing $T$. Between 230 and 160 K, the transfer of spectral weight from the in-gap states towards the peak opens a gap in the far-infrared $\sigma (\omega)$ around 150 cm$^{-1}$, in agreement with the results of Refs. \cite{Wang,Hwang}.
The $x$ = 0.57 sample (Fig.\ \ref{sigma}-b), the only one exhibiting a metal-like behavior, was extensively discussed in Ref. \onlinecite{Lupi}.

On the other hand, for $x \geq$ 0.75, the far-infrared $\sigma (\omega)$ looks like a combination of those for the charge-ordered  0.5 and the metallic 0.57 samples. At $x$ = 0.75 a broad absorption peak is centered around 200 cm$^{-1}$ at room temperature. Its intensity increases by lowering $T$ down to 30 K, whereas its characteristic frequency shifts to about 100 cm$^{-1}$. However, both at 12 and 8 K (the latter one is not shown in Fig.\ \ref{sigma}-c for sake of clarity) spectral weight is lost below 100 cm$^{-1}$ and transferred to higher frequencies. This effect is amplified both in the $x$ = 0.85 and 0.95 samples (Figs.\ \ref{sigma}-d and -e, respectively). Therein a strong FIP develops at nearly 100 cm$^{-1}$ below about 160 K and shifts to 200 cm$^{-1}$ below 30 K. One may notice that, below 22 K, a Spin-Density-Wave (SDW) instability has been reported for 0.7 $\le x \le $ 0.9. \cite{Foo,Sales}. The FIP peak is separated from the quasi-particle contribution, clearly visible for the $x$ = 0.75 and 0.95 samples below 50 and 80 cm$^{-1}$ respectively, by a minimum similar to that previously reported\cite{Wang2} at low $T$ in a sample with $x$ = 0.7.
Finally the nominally $x$ = 1 sample, at variance with the theoretical predictions\cite{Lee}, shows a metallic $\sigma (\omega)$ that, at low frequencies, increases by lowering $T$. However, this unexpected behavior will be explained in the next Section by the possible presence of a few Na defects in the nominally $x$ = 1 crystal.

\section{Discussion}

\subsection{Electronic  conductivity}

A FIP at finite energy associated with an optical gap, as in the present and previously reported \cite{Wang,Hwang}  $\sigma (\omega)$ of the $x$ = 0.5 sample, is usually observed in the infrared conductivity of charge-ordered materials.\cite{Dressel,Gruner}  According to the Charge-Density-Wave (CDW) theory, \cite{Dressel,Gruner}  the optical gap will measure the energy needed to excite one charge from its superlattice state (in Na$_x$CoO$_2$, the Co$^{3+}$-Co$^{4+}$ periodic structure) to E$_F$. In turn, a FIP in $\sigma (\omega)$, in the absence of full gap opening, reflects a coexistence of localized charges, not necessarily long-range ordered, and itinerant carriers. In this case, as it happens here for the samples with $x$ = 0.75 and 0.95, a minimum separating the Drude term from the localization peak may appear in the optical conductivity.

At $x$ = 0.5 the energy gap here is on the order of 150 cm$^{-1}$,  to be compared with the 800-1000 cm$^{-1}$ found in the charge-stripe state of the layered perovskite La$_{1.33}$Sr$_{0.67}$NiO$_4$. \cite{Paschke,Calvani2} This shows that charge ordering in Na$_{0.5}$CoO$_2$ affects smaller energy regions around E$_F$. However, the strong depression of $\sigma (\omega)$ below 150 cm$^{-1}$ is consistent with the presence of a gap over the entire Fermi surface.
The samples with $x$ = 0.75, 0.85, and 0.95 show a metallic dc conductivity  down to the lowest temperatures.\cite{Foo} However, as already mentioned, $\sigma (\omega)$ shows at $x$ = 0.75 (Fig.\ \ref{sigma}-c) a minimum around 50 cm$^{-1}$, which deepens by lowering $T$. A similar minimum (Fig.\ \ref{sigma}-e) appears around 80 cm$^{-1}$ in the $x$ = 0.95 sample. Both these minima separate a quasi-free particle term from a FIP at about 100 cm$^{-1}$, as observed in certain cuprates.\cite{Lupi00} This FIP is much softer than the corresponding one for the $x$ = 0.5 sample, but shows a similar temperature dependence.  Both at  $x$ = 0.75  and 0.95 the periodicity of magnetic correlations, measured by magnetic neutron scattering, \cite{Booth,Bayrakci}  should rule out the occurrence of long-range Co$^{3+}$-Co$^{4+}$ order in these samples. Therefore the soft FIP can be assigned to localized charges, either disordered or with short-range order, coexisting with itinerant carriers in the CoO$_2$ layers.

A clear shift to higher frequencies is observed in the FIP when $T$ is lowered from 30 K to 12 K (data at 8 K, virtually superimposed to those at 12 K, have not been reported): the FIP displaces to about 120, 200, and 200 cm$^{-1}$ for $x$ = 0.75 0.85, and 0.95, respectively. Those energies are close to that of the FIP in the 0.5 sample at 12 K, showing that further ordering takes place for the high-doping materials below 30 K. Between 30 and 12 K, moreover, a Drude-Lorentz fit to $\sigma(\omega)$ shows that the quasi-particle term looses about 60\% of its $SW$, both for $x$ = 0.75 and 0.95. At $x$ = 0.85, instead, the Drude term is not resolved down to 40 cm$^{-1}$. Since this material shows a metallic dc conductivity at low $T$,[2] one may infer that here the Drude term is weak and squeezed at very low frequencies.

The above strong effects on the low-energy charge dynamics at low $T$ show that the SDW transition at $T_{SDW}$ = 22 K \cite{Foo,Sales} further gaps the Fermi surface. From the FIP shift one can estimate for the charges an additional binding energy $E_{SDW}$ of about 100 cm$^{-1}$. This value is comparable with that measured for the SDW of Bechgaard salts (TMTSF)$_2$X (X=PF$_6$, ClO$_4$).\cite{Gruner} One thus obtains $E_{SDW}/k_BT_{SDW} \sim $ 6, a value consistent with a strong-coupling Spin-Density-Wave. Moreover, $E_{SDW}$ is of the same order as both the in-layer  and c-axis magnetic exchange interactions which characterize the A-type antiferromagnetic structure.\cite{Booth,Bayrakci}  Therefore the shift to higher energy of the FIP peak, below $T_{SDW}$, may be due to a strong interaction of the charge carriers with the in-layer ferromagnetically-ordered spins.
On the other hand, in the $x$ = 0.5 material, the FIP frequency monotonously decreases down to 8 K, indicating the absence of any SDW instability in this system. Therefore the charge ordering showed, above 30 K, by the 0.5 system or the charge localization in the $x$ = 0.75, 0.85 and 0.95 samples, must be due to a physical mechanism other than the SDW instability and may be correlated both to the pinning potential of the out-of-plane Na$^+$ ions and to the Hubbard repulsion between charge-carriers in the CoO$_2$ layers.

Finally, $\sigma (\omega)$ at $x$ = 1 unexpectedly does not show an insulating behavior, as instead predicted by band theory,\cite{Lee} but is poorly metallic and increases by lowering $T$ (Fig. 2-f). A FIP also appears at the lowest frequencies, between 200 and 160 K. These observations may be explained with a small number of Na$^+$ vacancies, which introduce holes in the otherwise full e$_{1g}$- a$_{g}$ electronic bands. Indeed the $SW$ of the $x$ =1 sample, obtained by integrating $\sigma (\omega)$ up to the pseudo-plasma edge ($\omega$ = 5000 cm$^{-1}$), is about 40 \% of the corresponding SW at $x$ = 0.95.  Assuming that the holes are only due to Na doping, the actual Na$^+$ content of the nominally $x$ = 1 sample can be estimated from the ratio $SW(x=1)/SW(x=0.95)$  to be 0.98.

\subsection{Phonon absorption}

Several authors have attributed the origin of charge ordering in Na$_x$CoO$_2$ to the out-of-plane ordering of the Na$^+$ ions \cite{Zand,Milne}. Indeed, their partially screened Coulomb potential may affect the mobility of the charge carriers in the CoO$_2$ layers. The "freezing" of the Na$^+$ lattice below a certain temperature would favor a pinning of the charge carriers. One may not expect any direct observation of Na$^+$ ordering in the infrared. However, as discussed in the Introduction, the "layer sliding" E$_{1u}$ phonon mode (well evident in the $\sigma (\omega)$ of all samples around 590 cm$^{-1}$, see Fig.\ \ref{sigma}) contains a small component of the Na$^+$ vibration parallel to the plane.\cite{Li} Therefore, the lifetime of this  mode may be sensitive to ordering phenomena at the Na$^+$ sites.

\begin{figure}
{\hbox{\psfig{figure=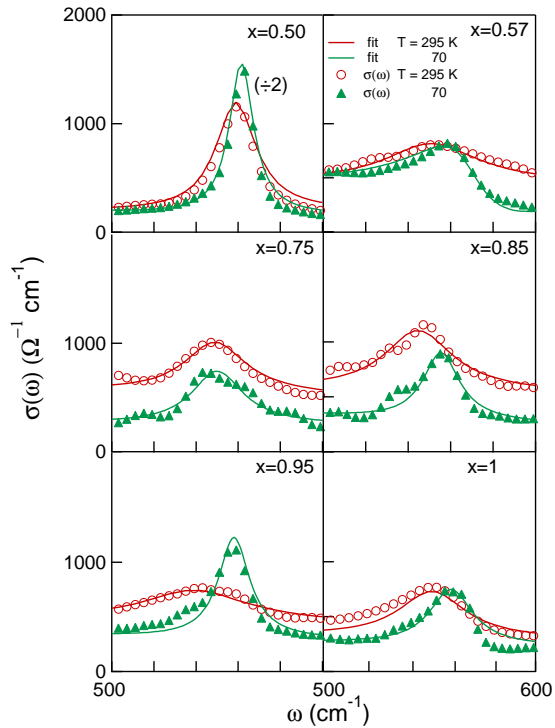,width=8cm}}} \caption{Real
part of the infrared conductivity corresponding to the E$_{1u}$
in-plane phonon at 295 (open symbols) and 70 K (full symbols).
Thin and thick solid lines are Drude-Lorentz fits to data  for
295 and 70 K, respectively (see text).} \label{phonon}
\end{figure}

Figure\ \ref{phonon} shows that this is indeed the case. Therein, $\sigma (\omega)$ is reported for the six samples at different $T$ in the E$_{1u}$ region (500 - 600 cm$^{-1}$), together with the Drude-Lorentz fit above mentioned. This latter provides the phonon parameters: intensity $S^2_{ph}$, frequency $\omega_{ph}$ and broadening $\Gamma_{ph}$. The results are rather accurate as the vibrational absorption, at any $x$, is well distinguished from the electronic background.

\begin{figure}
{\hbox{\psfig{figure=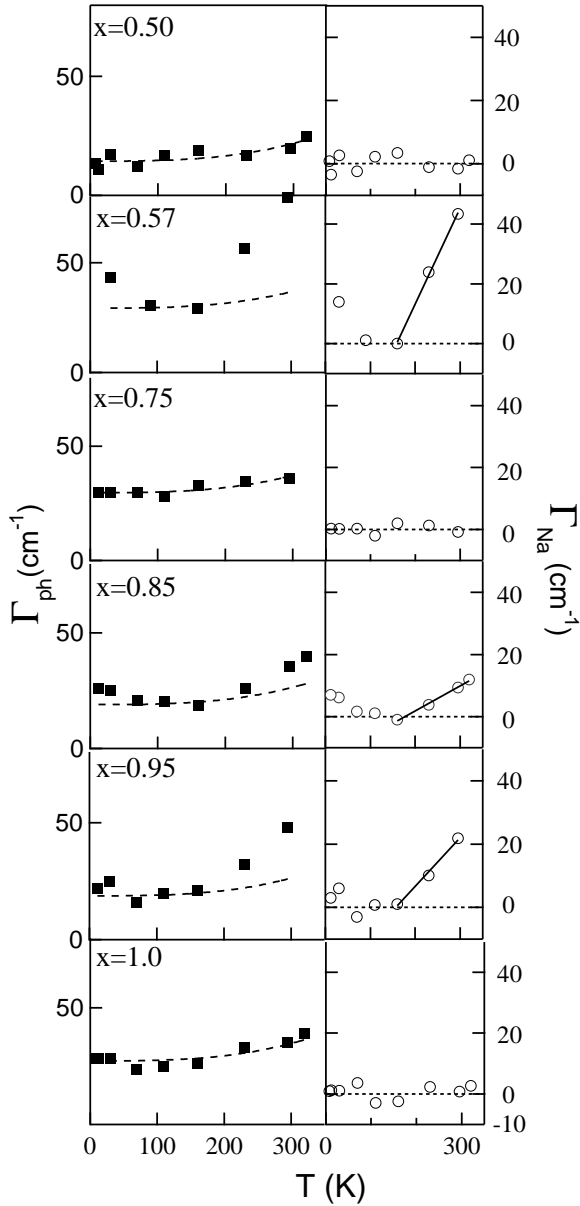,width=8cm}}} \caption{Temperature
dependence of the E$_{1u}$ phonon broadening $\Gamma_{ph}(T)$
(left panels, full symbols), and its fit (dashed lines) to Eq.\
\ref{fit} (see text). In the right panels an estimate of
$\Gamma_{Na}(T)$ (see Eq.\ \ref{sum}) to the phonon broadening
$\Gamma_{ph}(T)$ is shown for all samples. Here the solid line is
just a guide to the eye.} \label{gamma}
\end{figure}

The resulting line width $\Gamma_{ph}(T) $ is shown in Fig.\ \ref{gamma} (left panels) as a function of $T$ and $x$. One may notice that the $T$ dependence of $\Gamma_{ph}(T)$ is stronger for the incommensurate doping levels $x$ = 0.57, 0.85 and 0.95 than for its commensurate values. Several effects may influence the phonon broadening: disorder in the CoO$_2$ layers, inharmonic phonon-phonon interactions and disorder in the Na$^+$ lattice. The last two terms depend on $T$. Assuming that different decay channels are independent of each other, the phonon line width can be written as

\begin{equation}
\Gamma_{ph} (T) = \Gamma_{0} + \Gamma_{ph-ph} (T) + \Gamma_{Na}(T)\, .
\label{sum}
\end{equation}

\noindent
where $\Gamma_{0}$ is the contribution of structural disorder, both in the CoO$_2$ layers and in the Na$^+$ lattice, $\Gamma_{ph-ph}(T)$ is due to the inharmonic phonon-phonon interaction, and $\Gamma_{Na}(T)$ is related to thermal disorder and diffusion in the Na$^+$ lattice. Also in oxides, $\Gamma_{ph-ph}(T)$ is well described by the usual law $a(T/\Theta)^{3}$,\cite{Thomas}  where $a$ is a constant and $\Theta$ is the Debye temperature. If we assume, according to diffraction results, that in samples with commensurate doping ($x$ =0.5, 0.75 and 1.0) the Na$^+$ ions are already frozen at room temperature, $\Gamma_{Na}(T)$ = 0. Therefore, for those samples one has

\begin{equation}
\Gamma_{ph} (T) = \Gamma_{0} + a(T/\Theta)^{3} \, .
\label{fit}
\end{equation}

\noindent
Satisfactory fits to Eq.\ \ref{fit}, shown by the dashed lines in the left panels of Fig.\ \ref{gamma}, can be obtained for all samples with commensurate doping by varying $\Gamma_{0}$ only and using the same $a/\Theta^{3}$ = 3 x 10$^{-7}$ cm$^{-1}K^{-3}$. As $\Theta \simeq$ 380 K,\cite{Thomas} is not expected to depend appreciably on the Na content, one obtains an $x$-independent $a = 15\pm$4 cm$^{-1}$. The fit also gives $\Gamma_{0}$ = 25$\pm$5 cm$^{-1}$ for all samples, but for $x$ = 0.5 where $\Gamma_{0}$ = 10$\pm$2 cm$^{-1}$. Such a narrow line is then associated with the long-range Na order characteristic of the 0.5 sample. It is also consistent with that extracted from thermal conductivity measurements.\cite{Foo}

As shown in Fig.\ \ref{gamma}, at the incommensurate $x$ = 0.57, 0.85 and 0.95 one observes clear deviations from Eq. \ \ref{fit}, especially above 200 K. In those samples, Na$^+$ thermal disorder probably gives a contribution to $\Gamma_{Na}(T)$ also below 300 K. Moreover, one may notice that the increase of $\Gamma_{ph}$ below 100 K for $x$ = 0.57 (and probably for $x$ = 0.85 and 0.95) is due to Fano interaction between  this phonon mode and the quasi-particle continuum, as discussed in a previous paper.\cite{Lupi}
Since $\Gamma_{Na}(T)$ is not known, we cannot predict the temperature behavior of $\Gamma_{ph}(T)$ in the incommensurate systems. However, therein a freezing temperature for Na$^+$ diffusion can be determined by subtracting from the experimental $\Gamma_{ph} (T)$ that given by Eq.\ \ref{fit}, which holds for $\Gamma_{Na}(T) = 0$ (commensurate doping). The resulting  $\Gamma_{Na}(T)$  is shown by open symbols in the left panels of Fig.\ \ref{gamma}, where the solid line is only a guide to the eye.  Therein we also show the fit residual, which for commensurate samples is obviously zero. In Fig.\ \ref{gamma} $\Gamma_{Na}(T)$ vanishes around 150 K for all samples with incommensurate Na content. Such a freezing temperature is in qualitative agreement with the value (200 K) found by diffraction measurements in polycrystalline samples with $x >$ 0.7.\cite{Shi}

The present data also provide information on the interplay between the above detected Na$^+$ "freezing" and the signatures of charge localization in the infrared conductivity. The FIP already present at room temperature for commensurate doping ($x$ = 0.5 and x=0.75) develops between 230 and 160 K in the incommensurate systems with $x$ = 0.85 and 0.95. In the same samples and temperature range, as shown above, the phonon line width has an anomalous $T$-dependence due to increasing localization of the Na$^+$ ions. On the other hand, the FIP is not observed at all when the incommensurate doping is 0.57, even if Na$^+$ ions freeze at the same temperature as in the other two samples. It seems then that charge localization - and possibly short-range charge order - is driven by the freezing of out-of-plane Na$^+$ when the Na concentration is high and the charges density is low ($x$ = 0.85 and 0.95). This mechanism becomes much less efficient when the Na$^+$ concentration is low and the charge density is high ($x$ = 0.57).

\section{Conclusion}

In conclusion, we have carried out a systematic study of the optical conductivity of the Na$_x$CoO$_2$ system in the $x \geq 0.5$ part of its phase diagram. It is confirmed that charge ordering opens, at $x$ = 0.5, a far-infrared gap about 150 cm$^{-1}$ wide. As a result, a peak appears around 200 cm$^{-1}$, which corresponds to the lowest energy for exciting one charge out of the Co$^{3+}$-Co$^{4+}$ ordered ground state. At $x$ = 0.75, 0.85 and 0.95, the gap is partially filled by  Drude conductivity which is separated by a minimum from a far-infrared peak, centered at about 100 cm$^{-1}$ at 30 K. This FIP provides evidence that charge localization (or short-range charge order), competes with transport, since metallic conduction is observed at low $T$ also for $x >$ 0.5. Moreover, the magnetic ordering reported in the literature below 22 K and attributed to a Spin-Density-Wave instability induces, for incommensurate $x$ =  0.85 and 0.95, a strong renormalization of the charge dynamics in the CoO$_2$ planes. Indeed the far-infrared peak hardens dramatically below 30 K (by about 100 cm$^{-1}$) pointing towards a strong coupling of the charge and spin degrees of freedom. In contrast with these results, the $x$ = 0.57 sample shows a metallic conduction at each $T$, described by an anomalous-Drude behavior in the far-infrared $\sigma (\omega)$.

We have also shown that a correlation exists between the far-infrared peak and the out-of-plane localization of Na$^+$ ions, monitored by the E$_{1u}$ infrared-active phonon lifetime. In particular, the temperature where the FIP develops in the infrared conductivity is in reasonable agreement with the Na$^+$ freezing temperature, as extracted from the phonon line width. This mechanism, however, appears to be effective only for low charge density and high Na$^+$ concentration.

\end{document}